\documentclass[10pt]{article}

\usepackage[utf8x]{inputenc}

\usepackage{amsfonts}
\usepackage{graphicx}
\usepackage{color}
\usepackage{amsmath}
\usepackage{amssymb}
\usepackage{verbatim}
\usepackage{amsthm}

\usepackage{multirow}

\usepackage{setspace}

\usepackage{amsmath}
\usepackage{amssymb}

\usepackage{graphicx}

\usepackage{cite}

\usepackage{color}

\usepackage[labelfont=bf,labelsep=period,justification=raggedright]{caption}

\makeatletter
\renewcommand{\@biblabel}[1]{\quad#1.}
\makeatother

\date{}

\begin{document}

\begin{flushleft}
{\Large
\textbf{Beyond network structure: How heterogenous susceptibility modulates the spread of epidemics}
}
\\
Daniel Smilkov$^{1,2,\ast}$,
Cesar A. Hidalgo$^{1}$,
Ljupco Kocarev$^{2,3,4}$
\\
\bf{1} The MIT Media Lab, Massachusetts Institute of Technology, Cambridge, MA, USA \\
\bf{2} Macedonian Academy for Sciences and Arts, Skopje, Macedonia \\
\bf{3} BioCircuits Institute, University of California, San Diego, CA, USA\\
\bf{4} Faculty of Computer Science and Engineering, University ``Sv Kiril i Metodij'', Skopje, Macedonia  \\
$\ast$ E-mail: Corresponding author: smilkov@mit.edu
\end{flushleft}

\begin{abstract}
The compartmental models used to study epidemic spreading often assume the same susceptibility for all individuals, and are therefore, agnostic about the effects that differences in susceptibility can have on epidemic spreading. Here we show that--for the SIS model--differential susceptibility can make networks more vulnerable to the spread of diseases when the correlation between a node's degree and susceptibility are positive, and less vulnerable when this correlation is negative. Moreover, we show that networks become more likely to contain a pocket of infection when individuals are more likely to connect with others that have similar susceptibility (the network is segregated). These results show that the failure to include differential susceptibility to epidemic models can lead to a systematic over/under estimation of fundamental epidemic parameters when the structure of the networks is not independent from the susceptibility of the nodes or when there are correlations between the susceptibility of connected individuals.
\end{abstract}


The contact networks that underlie the spread of diseases, behaviors \cite{obesity} and ideas have heterogenous topologies \cite{barabasi, anderson-may, satorras2001-PRL}, but also, exhibit heterogeneity in the susceptibility of individuals \cite{hardie-aids2008,jli-plos2010, boom2011,friedman1972,nakamura1985}.
In recent years the compartmental models used to study the spread of behaviors have been generalized to include the effects of network topology, but not the effects of both topology and differences in the susceptibility of individuals \cite{satorras2001-PRL, satorras2002-PRE3, chakrabarti2003}. Generalizations including the network topology have revealed an intimate connection between the spectral properties of the contact network, and the basic reproductive number of infectious diseases, showing that for a network described by an arbitrary degree distribution the basic reproductive number of an infection ($R_0$), is proportional to the largest eigenvalue of the contact network's adjacency matrix \cite{chakrabarti2003}. For highly heterogenous networks, this eigenvalue is always larger than $1$ meaning that network heterogeneity can reduce and even eliminate the existence of an epidemic threshold.

Our understanding of the role of heterogeneous network topologies in epidemic spreading, however, has not been matched by a comparable development in our understanding of the role of heterogeneity in the susceptibility of individuals. Yet, differential susceptibility, defined as the variation in the susceptibility of individuals is as widespread as network heterogeneity. For example, genetic conditions are known to cause  heterogenous reactions to HIV \cite{hardie-aids2008,jli-plos2010}, H5N1 influenza \cite{boom2011}, and the Encephalomyocarditis virus, \cite{friedman1972} . Differential susceptibility can also be the result of differences in age as it has been shown in the case of Hantaan Virus in mice \cite{nakamura1985}. Other mechanisms leading to differential susceptibility include previous disease history, obesity, stress, history of drug abuse, physical trauma or differences in healthcare quality, which could emerge from discriminatory practices or individual self-selection. The biological prevalence of differential susceptibility, therefore, invites us to ask whether relaxing the assumptions of homogeneous susceptibility has consequences for the spread of epidemics that are tantamount to the relaxation of assumptions of homogeneity in the connectivity of the contact network.

The incorporation of differential susceptibility into epidemic models, however, also introduces a new dimension to epidemic modeling, since there are multiple ways for individuals with differences in susceptibility to be arranged in a network. For instance, the mixing patterns and segregation of populations \cite{newman-PRE} imply that differential susceptibility can be structured through non-trivial correlations. Examples here include schools, nursing homes and hospitals, where  children, senior citizens and patients, who can be more susceptible to diseases, spend more time together. These mixing patterns imply that a complete study of differential susceptibility should consider not only variations in the susceptibility of individuals, but also the  correlations between the susceptibility of individuals and the positions these occupy in a network.

Understanding how epidemic spreading is affected by differential susceptibility can affect a number of policy decisions, since epidemic models do not only inform the spread of infectious diseases, but also the spread of behaviors \cite{complex-contagion}, such as smoking \cite{smoking}; health conditions, such as obesity \cite{obesity}; and digital threats, such as computer and mobile phone viruses \cite{mobility}. Here, however, we show that the failure to consider differences in the susceptibility of individuals can lead to over- or underestimate a network's vulnerability to epidemic spreading.

In this paper we solve the SIS epidemic model for a contact network with arbitrary network topology and differential susceptibility and show that its basic reproductive number $R_0$ is proportional to the maximal eigenvalue of a topology-infection matrix that combines information on the topology of the network and the susceptibility of individuals. Using mean-field theory, we look at individual level correlations between the susceptibility and the degree or connectivity of individuals and show that positive correlations between susceptibility and degree makes the network more vulnerable to epidemics (increasing $R_0$), whereas negative correlations make the network less vulnerable (decrease $R_0$). Finally we look at segregation by studying the consequences of having individuals connected to other individuals with similar characteristics and show that segregation significantly increases the vulnerability of the network to disease -- causing $R_0$ to increase. To conclude, we illustrate the strong effects of segregation dynamics on $R_0$ by running a variant of Schelling's segregation process \cite{schelling} on a real-world contact network obtained from face-to-face proximity between students and teachers. This shows then even a mild level of segregation can drastically increase the critical reproductive number in a network where individuals differ in their susceptibility.

\section*{Results}
\subsection*{The basic reproductive number $R_0$ in the SIS model}
We begin by summarizing the main results for the SIS model in heterogenous networks without differential susceptibility. This will help us introduce the methodology and notation that we will use later, and will also help us compare known results with those obtained for networks when differential susceptibility is present.

In the SIS model individuals can exist in either of two possible states: ``healthy'' or ``infected''. Healthy individuals are infected when they come into contact with an infected individual with probability $\beta$. Infected individuals, on the other hand, become once again susceptible with a recovery probability $\delta$. When each individual is in contact with $k$ others, the basic reproductive number of a homogenous network takes the form:
\begin{equation}
\label{r0-mixed}
R_0^{h}=\frac{k\beta}{\delta}
\end{equation}
where the superscript $h$ is used to indicate a homogenous network. $R_0^{h}$ can be interpreted as the average number of new infections that an infected individual generates during his infective period in a fully susceptible population. $R_0$ is the quintessential epidemiological parameter, since the infection can only spread when an infected individual gives rise to one or more new infected ($R_0>1$). Because of this, a central question in epidemiology is under what conditions $R_0$ becomes greater than 1.

In degree heterogenous networks with no degree-degree correlations and a degree distribution given by $P(k)$ the basic reproductive number $R_0$ generalizes to  \cite{anderson-may, satorras2001-PRL}
\begin{equation}
\label{r0-uncorrelated}
R_0^{unc} = \frac{\beta}{\delta}\frac{\left<k^2\right>}{\left<k \right>} = R_0^{h}\left( 1+ \left[\frac{\sigma_k}{\left<k \right>}\right]^2 \right)
\end{equation}
where the superscript $unc$ stands for uncorrelated networks, $\left<k\right>$ is the average degree of the contact network, $\left<k^2\right>$ is the average of the degrees squared and $\sigma_k$ is the standard deviation of the degree distribution. Note that when there is no heterogeneity, equation (\ref{r0-uncorrelated}) reduces to equation (\ref{r0-mixed}), otherwise $R_0^{unc} > R_0^{h}$ since $\left<k^2\right> > \left<k\right>^2$. In fact, for highly heterogenous networks, where the degree distribution follows a power-law $P(k) \propto k^{-\alpha}$ with $\alpha<3$, $\sigma_k$ (and therefore $R_0$) grows with the network size, implying that the epidemic threshold vanishes for infinitely large networks ($R_0$ is always larger than 1). It is worth noting that this result was obtained using the heterogenous mean-field (HMF) theory, which neglects both dynamical and topological correlations. In HMF theory, the actual quenched structure of the network given by
its adjacency matrix $A_{ij}$ is replaced by an annealed version, in which edges are constantly rewired at a rate much faster than that of the epidemics, while preserving the degree distribution $P(k)$ \cite{langevin}.

In most real world networks nodes are not connected randomly to other nodes, and this effect is characterized by what is known as non-trivial mixing patterns \cite{newman-PRE}. In the case of degree heterogeneity these mixing patterns are captured by the conditional probability that a link starting at a node with degree $k$ will end at a node with degree $k'$ ($P(k'|k)$). In this case, the basic reproductive number takes the form \cite{satorras2002-PRE3}
\begin{equation}
R_0^{corr} = \frac{\beta}{\delta} \lambda_{1,C}
\end{equation}
where $\lambda_{1,C}$ is the largest eigenvalue of the degree mixing matrix $C_{kk'} = kP(k'|k)$.

While HMF theory assumes annealed networks, its validity for real (quenched) networks is limited \cite{nature-epidemic-threshold}. Quenched networks are static, given by the adjacency matrix $A$, where $a_{ij}=1$ if there is a link connecting node $i$ to node $j$ and $0$ otherwise. A significant improvement over the HMF theory is given by the quenched mean-field (QMF) theory where it has been shown that the basic reproductive number takes the form \cite{chakrabarti2003}:
\begin{equation}
\label{r0-qmf}
R_0^{QMF} = \frac{\beta}{\delta}\lambda_{1,A}
\end{equation}
where $\lambda_{1,A}$ is the largest eigenvalue of the matrix $A$.

\subsection*{Differential susceptibility}

Next, we introduce differential susceptibility by assuming that the susceptibility probability $\beta_i$ is different for each node $i$ in the network. This generalization introduces an important conceptual difference in our interpretation of susceptibility. In models of homogeneous susceptibility $\beta$ is interpreted as the probability that an infected individual passes on the disease to a susceptible individual, and hence, $R_0$ is interpreted as the average number of infected individuals begot by each infected individual in a fully susceptible population. This assumption of directionality is needed because one individual can infect many others, but cannot be infected by more than one individual. When the fraction of infected individuals is small, however, $\beta$ can also be interpreted as the probability that an individual gets infected, since the probability that a susceptible individual is in contact with multiple infected individuals is negligible. Biologically speaking, thinking of susceptibility in terms of the probability that a susceptible individual gets infected makes more sense than thinking of susceptibility in terms of the probability that an infected individual passes on the disease. Because of this biological consideration, we flip the standard assumption and pin the variation in susceptibility to the susceptible individual. We note this approximation is valid only when the overall level of infection is low, which is the limit in which we derive our results. 

After taking this consideration into account we proceed by following Wang et al. \cite{chakrabarti2003} and note that for an arbitrary network topology given by an irreducible non-negative adjacency matrix $A = [ a_{ij} ]_{N \times N}$, the evolution of the SIS model can be written following as:
\begin{equation}
\label{evolutionSIS} p_i(t+1) = (1 - p_i(t)) f_i(t) + (1-\delta) p_i(t)
\end{equation}
where $p_i(t)$ is the expected probability that node $i$ will be infected at time $t$, and $f_i(t)$ is the probability that node $i$ receives the infection from at least one of its infected neighbors at time $t$. The probability $f_i(t)$ has the form:
\begin{equation}
\label{fi} f_i(t) = 1 - \prod_{j=1}^N (1 - \beta_i a_{ij} p_j(t)).
\end{equation}
Equation (\ref{evolutionSIS}) is a non-linear dynamical system in which the case of no infection in the network ($p_i = 0, i \in \{1, \ldots, N \}$) represents a fixed point of the system that becomes unstable when $R_0>1$. To test for asymptotic stability, we linearize the system (\ref{evolutionSIS}) around $\mathbf{p}=0$ by removing all higher-order terms of $p_i(t)$:
\begin{equation}
\label{evolutionSIS-linear} p_i(t+1) = (1-\delta)p_i(t) + \sum_{j=1}^N \beta_i a_{ij} p_j(t) = \sum_{j=1}^N m_{ij} p_j(t).
\end{equation}
We note that here we have assumed no differences in the recoverability rate of individuals (i.e. $delta$ is constant). Next, we express (\ref{evolutionSIS-linear}) in matrix form as:
\begin{equation}
\mathbf{p}(t+1) = M \mathbf{p}(t)
\end{equation}
where $M=[m_{ij}], m_{ij}=\beta_i a_{ij} + \Delta_{ij}(1-\delta)$ and $\Delta_{ij}$ is the Kronecker delta. Thus, one gets that the state with no disease is asymptotically stable if the largest eigenvalue of $M$,  $\lambda_{1,M}<1$.

We note that when the origin is asymptotically stable, it is also globally stable, since
\begin{equation}
(1 - p_i(t)) f_i(t) \leq f_i(t) \leq \sum_{j=1}^N \beta_i a_{ij} p_j(t)
\end{equation}
where the second inequality follows directly from the Weierstrass product inequality. Using this, we see that the system satisfies the inequality
\begin{equation}
\mathbf{p}(t) \leq M \mathbf{p}(t-1) \leq M^t \mathbf{p}(0).
\end{equation}
Thus, when $\lambda_{1,M}<1$, the infection will die out exponentially fast with a rate determined by $\lambda_{1,M}$.

\subsection*{Solutions with incomplete information}
The connection between the largest eigenvalue of $M$ ($\lambda_{1,M}$) and the epidemic threshold represents a solution of the system that has little practical use in absence of complete information about the network topology and the susceptibility of individuals. For the model to be of practical use we need to estimate $\lambda_{1,M}$ when there is incomplete information about the topology of the network and/or the distributions of susceptibility probabilities of individuals. First, we note that the matrix $M = R + (1-\delta)I$ where $R = [r_{ij}], r_{ij} = \beta_i a_{ij}$ and $I$ is the identity matrix. Therefore, we can define the threshold at which epidemics begin to spread through the largest eigenvalue of $R$ and generalize $R_0$ in equation (\ref{r0-qmf}) to:
\begin{equation}
\label{r0-qmf-general}
R_0^{QMF} = \lambda_{1,R} + 1 - \delta.
\end{equation}

To estimate $\lambda_{1,R}$ when there is incomplete information about the system, we assume that $r_{ij}$ is a random variable following an arbitrary distribution and use a mean-field approximation $r_{ij} \approx \left< r_{ij}\right>$ where $\left< r_{ij}\right>$ is the expected value of $r_{ij}$ over all possible network realizations.
In a network where the susceptibility $\beta_i$ is assigned independently of the topology, we have that $\left< r_{ij} | a_{ij}\right> = \left< \beta | a_{ij} \right> a_{ij} = \left< \beta \right> a_{ij} $ where $\left< x | y \right>$ is the expected value of $x$ given $y$. Then we obtain the basic reproductive number
\begin{equation}
\label{threshold-general}
R_0^{ind} =  \left< \beta \right> \lambda_{1,A} + 1 - \delta.
\end{equation}

In the case of uncorrelated networks with heterogenous degrees, we relax $\left< r_{ij}\right>$ to the expected number of links between nodes $i$ and $j$, which is proportional to the product of the degrees of $i$ and $j$. Keeping arbitrary susceptibility probabilities and degrees, we have
\begin{equation}
r_{ij} \approx \left< r_{ij}\right> = \beta_i \frac{k_i k_j}{ N \left< k \right>}.
\end{equation}
Since $R$ is an irreducible matrix, the eigenvector $\mathbf{v}$ associated with the maximal eigenvalue $\lambda_{1,R}$ is strictly positive.  Additionally, there are no other positive eigenvectors except positive multiples of $\mathbf{v}$. In this case, using the positive eigenvector $\mathbf{v}=[v_i]$ with $v_i = k_i \beta_i $, one immediately obtains the maximal eigenvalue, and therefore the basic reproductive number
\begin{equation}
\label{eq:local_corr}
R_0^{unc} = \frac{\left< \beta k^2 \right>}{\left< k \right>} + 1 - \delta = \frac{\left< \beta \right> \left< k^2 \right>}{\left< k \right>} + \frac{\rho \sigma_{\beta} \sigma_{k^2}}{\left< k \right>} + 1 - \delta.
\end{equation}
where $-1\leq \rho \leq 1$ is the Pearson correlation coefficient between the susceptibility $\beta_i$ and the square of the degree $k_i$ of individuals, $\left<\beta \right>$ and $\left< k^2 \right>$ are their respective averages, and $\sigma_{\beta}$ and $\sigma_{ k^2 }$ are their respective standard deviations. Looking at equation (\ref{eq:local_corr}), we point out that even small correlations between the susceptibility and the degree can lead to significant over- or underestimation of $R_0$ when the variation in connectivity, as measured by $\sigma_{ k^2 }$, is large compared to the average connectivity $\left< k \right>$ , which is the for networks following a heterogenous degree distribution. 


\subsubsection*{Networks with non-trivial mixing patterns}
We now focus on link level correlations, where the tendency of individuals to connect to other individuals with similar characteristics leads to non-trivial mixing patterns. We note that the same mathematical procedure can be used for the analysis regardless of whether we have correlations in susceptibility and constant recovery probability or vice versa. We consider link level correlations in the degree and the susceptibility of nodes. Then we approximate $r_{ij}$ with $\left< r_{ij}\right>$, by using the expected number of links from node with degree $k_i$ and susceptibility $\beta_i$ to node with degree $k_j$ and susceptibility $\beta_j$. The expected number of links are proportional to the two-point conditional probability $P(k', \beta' | k, \beta)$:
\begin{equation}
\label{r-definition}
r_{ij} \approx \left< r_{ij}\right> = \frac{k_i \beta_i P(k_j, \beta_j | k_i, \beta_i)}{N P(k_j, \beta_j)}.
\end{equation}
To find the maximal eigenvalue of the matrix $R$ as defined by equation (\ref{r-definition}) we need to find a positive vector $\mathbf{v} = [v_i]$ such that
\begin{equation}
\label{rewrite1}
v_i \lambda_{1,R} = \frac{k_i \beta_i }{ N} \sum_j \frac{P(k_j, \beta_j | k_i, \beta_i)}{ P(k_j, \beta_j)} v_j.
\end{equation}
holds for all $i$. Using some algebraic manipulations, we can rewrite equation (\ref{rewrite1}):
\begin{equation}
v_i \lambda_{1,R} = \frac{k_i \beta_i}{ N} \sum_{k', \beta'}  \left[ \sum_{j, k_j = k', \beta_j = \beta'} \frac{P(k_j, \beta_j | k_i, \beta_i)}{P(k_j,\beta_j)} v_j \right]  = \frac{k_i \beta_i}{ N} \sum_{k', \beta'} \frac{P(k', \beta' | k_i, \beta_i)}{P(k',\beta')} \sum_{j, k_j = k', \beta_j = \beta'} v_j
\end{equation}
which holds for every $i$. Writing $\sum_{j, k_j = k', \beta_j = \beta'} v_j = v_{k', \beta'}$ and summing these $N$ equations over $i$ where $k_i=k$ and $\beta_i = \beta$, we have
\begin{equation}
\label{rewrite3}
v_{k, \beta}  \lambda_{1,R} = \sum_{i, k_i=k, \beta_i = \beta} \frac{k_i \beta_i}{ N} \sum_{k', \beta'} \frac{P(k', \beta' | k_i, \beta_i)}{P(k',\beta')} v_{k', \beta'} = k \beta P(k, \beta) \sum_{k', \beta'} \frac{P(k', \beta' | k, \beta)}{P(k', \beta')} v_{k', \beta'} 
\end{equation}
which holds for every combination of $k$ and $\beta$. We can simplify equation (\ref{rewrite3}) further by writing $v_{k, \beta} = P(k, \beta) \hat{v}_{k, \beta}$ :
\begin{equation}
\label{rewrite4}
\hat{v}_{k, \beta} \lambda_{1,R} = \sum_{k', \beta'} k \beta P(k', \beta' | k, \beta) \hat{v}_{k', \beta'} = \lambda_{1,D} \hat{v}_{k, \beta}.
\end{equation}
where $\lambda_{1,D}$ is the maximal eigenvalue of the matrix $D_{\{k,\beta\},\{k', \beta'\}} = k \beta P(k', \beta' | k, \beta)$. In other words, the matrices $R$ and $D$ share the same eigenvalue. So, we will use $\lambda_{1,D}$ as an approximation of the actual $R_0$ when there is limited information in the network:
\begin{equation}
\label{r0-limited}
R_0 = \lambda_{1,R} + 1 - \delta \approx \lambda_{1,D} + 1 - \delta.
\end{equation}
Assuming d=$| \{k,\beta\} |$ is the number of different combinations of degree ($k$) and susceptibility ($\beta$) that a node can have in the network, we have compressed the entire information about the system (network topology and the susceptibilities of nodes) into a coarsened $d \times d$ matrix. In fact, we can choose $d$ depending on how much information we have about the network. For example, we can assign each individual to one of $5$ degree classes, $k=1,2,4,8,16$ and to one of $3$ susceptibility classes, $\beta=\beta_{low}, \beta_{avg}, \beta_{high}$ corresponding to low, average and high susceptibility respectively. We then only need to estimate the mixing patterns $P(k', \beta' | k, \beta)$ between the $15$ classes of individuals, as opposed to knowing every entry in the matrix $R$. The more classes we have, the more mixing patterns we have to estimate and the closer $\lambda_{1,D}$ will be to the actual $\lambda_{1,R}$.

Moreover, if we assume independence between the degree of a node and its susceptibility, i.e. $P(k', \beta' | k, \beta) = P(k'| k)P(\beta' | \beta)$ and choose $v_{k, \beta} = v^1_{k} v^2_{\beta}$ where $\mathbf{v}^1 = [v^1_k]$ and $\mathbf{v}^2 = [v^2_\beta]$ are the positive eigenvectors corresponding to the maximal eigenvalues $\lambda_{1,C}$ and $\lambda_{1,B}$ of the matrices $C_{kk'} = kP(k'|k)$ and $B_{\beta \beta'} = \beta P(\beta' | \beta)$ respectively, we can simplify equation (\ref{rewrite4}):
\begin{equation}
v^1_{k} v^2_{\beta}  \lambda_{1,R} = \sum_{k'} k P(k' | k) v^1_{k'} \sum_{\beta'} \beta P(\beta' | \beta) v^2_{\beta'} = \lambda_{1,C} \lambda_{1,B} v^1_{k} v^2_{\beta} 
\end{equation}
which gives $\lambda_{1,R} = \lambda_{1,C} \lambda_{1,B} $ and solves the system of equations. 
Note that when there is no degree mixing, $\lambda_{1,C}=\left< k^2 \right> / \left< k \right>$. On the other hand, when there are no mixing patterns in susceptibility, $\lambda_{1,B} = \left< \beta \right>$.

Equation (\ref{rewrite4}) allows us to estimate the basic reproductive number at different granularity depending on how much information we have about the network topology and the distribution of susceptibilities.  However, from the scientific perspective it is still unclear how the different mixing patterns impact the spread of diseases. To push our understanding of the effect of differential susceptibility further we study the effects of segregation in the spread of diseases.

\subsection*{Segregation}
Since the contact networks that underlie the spread of a disease are not only heterogeneous in terms of degree and susceptibility, but also segregated, we next proceed to solve the model for the cases where individuals are more likely to connect with others a similar level of susceptibility. As we will see, segregation can have an impact in the spreading of a disease that goes beyond the effects of differential susceptibility.

To understand the effect of segregation, we focus on the case where the degree of a node is independent of its susceptibility, $\lambda_{1,R} = \lambda_{1,C} \lambda_{1,B} $, and look respectively, at the limiting cases when there is no segregation and maximal segregation. Note that in the case of no segregation $P(\beta' | \beta) = P(\beta')$ and the largest eigenvalue of $B$ is $\lambda_{1,B}=\left< \beta \right>$ (with eigenvector $\mathbf{v}$ , $v_\beta=\beta$), hence the basic reproductive number takes the form
\begin{equation}
R_0 =  \left< \beta \right> \lambda_{1,C} + 1 - \delta.
\end{equation}
On the other hand, in the case of maximal segregation, where nodes only share links with nodes with the same susceptibility, $B$ is diagonal with $\beta$'s as the diagonal elements. In this case, $\lambda_{1,B}$ is equal to the largest susceptibility in the system  $\beta_{max}$. Thus, in the case of maximal segregation, the basic reproductive number becomes
\begin{equation}
R_0 = \beta_{max} \lambda_{1,C} + 1 - \delta.
\end{equation}
We note that in this extreme case nodes do not share links outside their susceptibility class, so the network is made of disconnected components. Hence, $R_0>1$ implies a persistent infection in at least one of these components. We do want to stress that in general $R_0>1$ does not guarantee a macroscopic outbreak in the network, but the existence of a highly vulnerable pocket, or subgraph, where $R_0$ is above $1$. In this case, the epidemic will be persistent, but will most likely  remain contained.

To obtain general bounds for $R_0^{corr} = \lambda_{1,C} \lambda_{1,B} + 1 - \delta$, we use the Collatz-Wielandt formula \cite{matrix-analysis}, which states that the Perron root of a matrix $A$ is given by $r=\max_{\mathbf{v} \in \Omega} f(\mathbf{v})$, where
\begin{equation}
f (\mathbf{v}) = \min_{i, v_i \neq 0} \frac{1}{v_i} \sum_{j} a_{ij} v_j \mbox{ and } \Omega = \{ \mathbf{v} | \mathbf{v} \geq 0 \mbox{ with } \mathbf{v} \neq 0 \}.
\end{equation}
Similarly, the min-max version states that $r=\min_{\mathbf{v} \in \Omega} f(\mathbf{v})$ where
\begin{equation}
f (\mathbf{v}) = \max_{i, v_i \neq 0}\frac{1}{v_i} \sum_{j} a_{ij} v_j .
\end{equation}
We rewrite the min-max and max-min versions of the Collatz-Wielandt formula into a form that we will use in the rest of the paper
\begin{equation}
\label{eq:collatz}
\min_{i, v_i \neq 0} \frac{1}{v_i} \sum_{j} a_{ij} v_j \leq r \leq \max_{i, v_i \neq 0} \frac{1}{v_i} \sum_{j} a_{ij} v_j
\end{equation}
which holds for all non-negative non-zero vectors $\mathbf{v}$. Assuming the matrix $B$ is irreducible, the Perron root of $B$ coincides with the largest eigenvalue $\lambda_{1,B}$. Thus, using equation (\ref{eq:collatz}), we can bound $\lambda_{1,B}$
\begin{equation}
\label{bound1}
\min_{\beta, v_{\beta} \neq 0} \frac{\beta}{ v_{\beta}} \sum_{\beta'} P(\beta' | \beta)v_{\beta'} \leq \lambda_{1,B} \leq \max_{\beta, v_{\beta} \neq 0} \frac{\beta}{ v_{\beta}} \sum_{\beta'} P(\beta' | \beta)v_{\beta'}
\end{equation}
where $\mathbf{v} = [v_\beta]$ can be any non-negative vector. Choosing the vector $\mathbf{v}$ such that $v_{\beta} = \beta$ in equation (\ref{bound1}), we immediately obtain the bound:
\begin{equation}
\lambda_{1,C} \min_{\beta'} \left< \beta | \beta' \right> + 1 - \delta  \leq R_0 \leq \lambda_{1,C} \max_{\beta'} \left< \beta | \beta' \right> + 1 - \delta
\end{equation}
where $\left< \beta | \beta' \right>$ denotes the average susceptibility of the individuals connected to an individual with susceptibility $\beta'$. These bounds provide useful information to understand the potential impact of a disease under limited knowledge about the network topology. We note that weak segregation limits the variation of $\left< \beta | \beta' \right>$ across the different susceptibility classes bounding $R_0$ in a small region around $\left< \beta \right>$. Strong segregation on the other hand, increases the gap between the bounds and our uncertainty for $R_0$.

\begin{figure}[!ht]
\begin{center}
\includegraphics[scale=0.6]{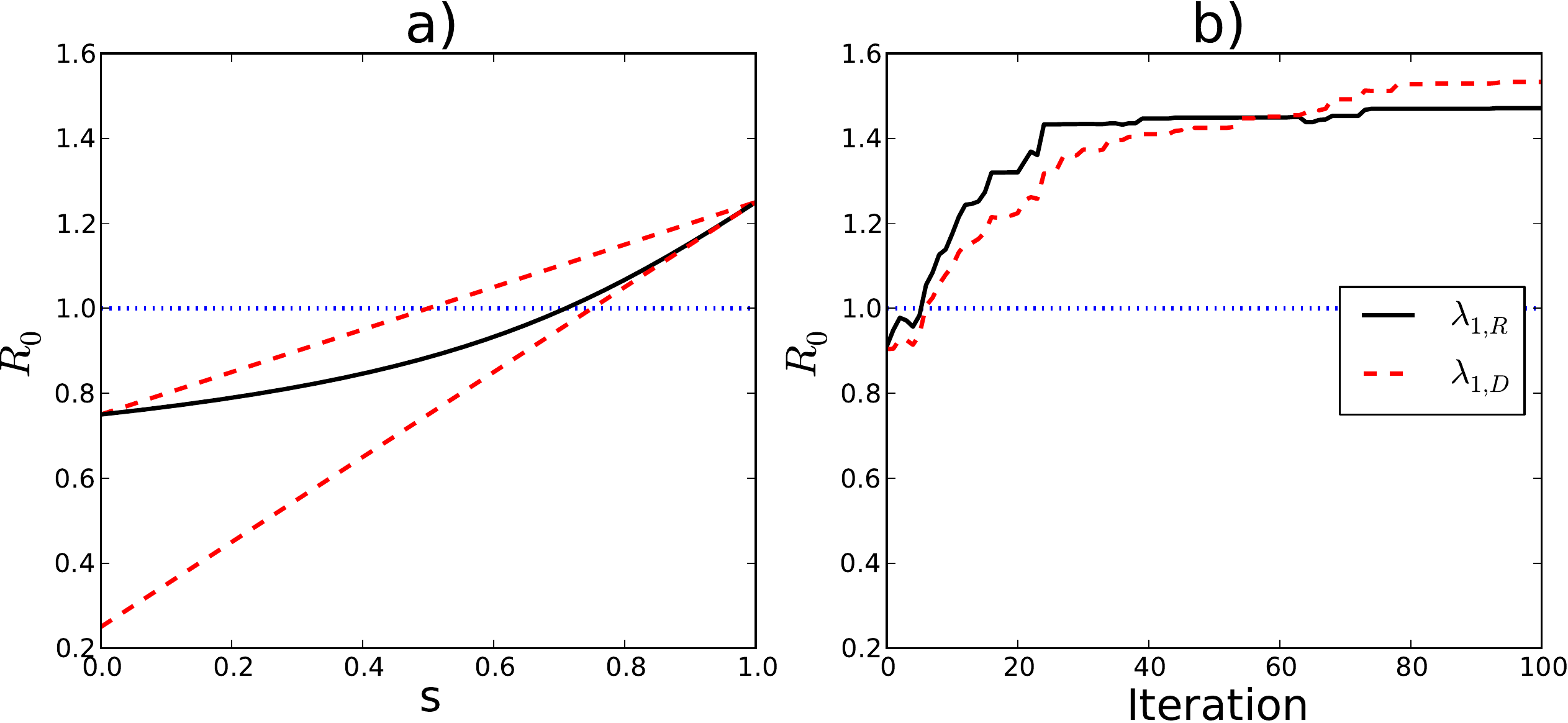}
\end{center}
\caption{ {\bf Impact of segregation on $R_0$}. a) $R_0= \lambda_{1,B}\left<k\right> + 1- \delta$ as a function of the amount of segregation $s$ (solid line) along with the bounds given by equation (\ref{segbounds}) (dashed lines) with $N_{\beta}=5$ susceptibility classes uniformly spaced between $[0.005, 0.025]$ with $\left<\beta \right>=0.015$. Additionally, to compute an actual value for $R_0$ a degree $k=25$ and $\delta=0.5$ was assumed. b) $R_0=\lambda_{1, R}$ (solid line) and the approximation $R_0 \approx \lambda_{1,D}$ (dashed line) for the face-to-face proximity school network for different iterations of the segregation process.}
\label{fig:segmodel}	
\end{figure}

To extend our intuition further we consider a simple network model with tunable segregation. Here, each node has the same degree $\left<k \right>$ and $P(\beta' | \beta) = (1-s)/N_{\beta}$ and $P(\beta | \beta) = (1-s)/N_{\beta} + s$ where $N_{\beta}$ is the number of different susceptibility classes and $s \in [0,1]$ models the segregation in the network. When $s=0$,  $P(\beta' | \beta)=1/N_{\beta} = P(\beta | \beta)$ and the chances that a node will connect to others is independent of the susceptibility $\beta$. When, $s=1$, segregation is maximal ($P(\beta|\beta)=1$ and $P(\beta' | \beta)=0$) and nodes share links only with others that have the same susceptibility. In this model, $R_0 = \lambda_{1,C} \lambda_{1,B} + 1 - \delta = \left< k \right>\lambda_{1,B} + 1- \delta $ since all nodes have the same degree $\left<k \right>$. To derive bounds for $R_0$ in this segregation model, we note that $P(\beta' | \beta) = (1-s)/N_{\beta} + \Delta_{\beta \beta'}s$ where $\Delta_{\beta \beta'}$ is the Kronecker delta. Putting this into equation (\ref{bound1}) we have
\begin{equation}
\lambda_{1,B} \leq \max_{\beta, v_{\beta} \neq 0} \frac{\beta(1-s)}{N_\beta v_\beta} \sum_{\beta'} v_{\beta'} + s\beta.
\end{equation}
Then choosing $v_\beta = \beta P(\beta)$ and using the fact that $P(\beta) = 1 / N_\beta$ we have
\begin{equation}
\lambda_{1,B} \leq \max_{\beta} (1-s)\left< \beta \right> + s\beta = (1-s)\left< \beta \right> + s\beta_{max}.
\end{equation}
On the other hand, the lower bound for $\lambda_{1,B}$ has the form
\begin{equation}
\label{eq:mintau}
\min_{\beta, v_{\beta} \neq 0} \frac{\beta(1-s)}{N_\beta v_\beta} \sum_{\beta'} v_{\beta'} + s\beta \leq \lambda_{1,B}
\end{equation}
where again $\mathbf{v} = [v_\beta]$ can be any non-negative vector. Note that if we choose $v_\beta = 1$ if $\beta=\beta_{max}$ and $0$ otherwise, we get the lower bound
\begin{equation}
\beta_{max} \left[ (1-s)/N_{\beta} + s \right] \leq \lambda_{1,B}.
\end{equation}
Thus, we obtain the bounds for the basic reproductive number:
\begin{equation}
\label{segbounds}
\left<k \right> \left[\frac{1-s}{N_{\beta}} + s\right]\beta_{max} + 1 - \delta \leq R_0 \leq \left<k \right> \left[(1-s)\left< \beta \right> + s \beta_{max} \right] + 1 -\delta.
\end{equation}
Fig. ~\ref{fig:segmodel}a shows the basic reproductive number $R_0$ as a function of the amount of segregation $s$ (solid line) along with the bounds given by equation (\ref{segbounds}) (dashed lines). We observe that as the amount of segregation increases, the network develops a highly susceptible pocket and $R_0$ tends to $\left< k \right> \beta_{max} + 1 - \delta$. This increase in vulnerability comes from highly susceptible groups of individuals that provide a stable pocket supporting the infection. The results show that segregation makes the network of individuals only as strong as its weakest subgroup. This suggests an immunization strategy that seeks to identify and target clusters of highly susceptible individuals, instead of groups of highly susceptible individuals that are connected to others that are less susceptible.

To illustrate the effects of segregation, we look at a real-world contact network of face-to-face proximity between students and teachers in a primary school \cite{primary-school}.  Links between A and B denote the cumulative time spent by A and B in face-to-face proximity, over one day. For simplicity, we convert this daily network to an unweighted network by assigning a link between the pair of nodes that have spent more than 2 minutes of cumulative time in face-to-face proximity. After discretizing the links, we keep the largest connected component which consists of $225$ nodes. We then run a variation of the Schelling's segregation process \cite{schelling} where initially each node is randomly assigned to one of two susceptibility classes. We use $\delta = 0.5$ and assign high susceptibility nodes $\beta_{high}=\beta_1$ and low susceptibility nodes with $\beta=\beta_{1} / 10$, thus having only one parameter in the network, $\beta_1$. To segregate this initial network, we assign to each node $i$ a potential energy $\left| \beta_i - \left< \beta \right>_i \right|$ where $\left< \beta \right>_i$ is the average susceptibility of $i$'s neighboring nodes. At each iteration, we swap a random pair of nodes if this decreases the total potential energy in the network. With each new iteration, we conserve the distribution of total susceptibility in the network but increase the level of segregation. Fig.~\ref{fig:segmodel}b shows the increase of $R_0=\lambda_{1, R}$ (solid line) with iterations of Schelling's segregation process. The dashed line on the other hand shows the approximation $R_0 \approx \lambda_{1,D}$ when there is limited information in the network. The matrix $D$ is constructed by coarse-graining the network into 7 degree classes $k=\{1,2,4,8,16,32,64\}$ where each node is assigned to its nearest degree class and two susceptibility classes $\beta=\{\beta_{1}, \beta_{1}/10 \}$. It is worth mentioning that we compute $\lambda_{1,D}$ by estimating only $14^2$ entries of the matrix $D$, as opposed to the $225^2$ entries of the matrix $R$, which is a significant reduction in the amount of information we need about the system.

\begin{figure*}[!ht]
\begin{center} 
\includegraphics[scale=0.60]{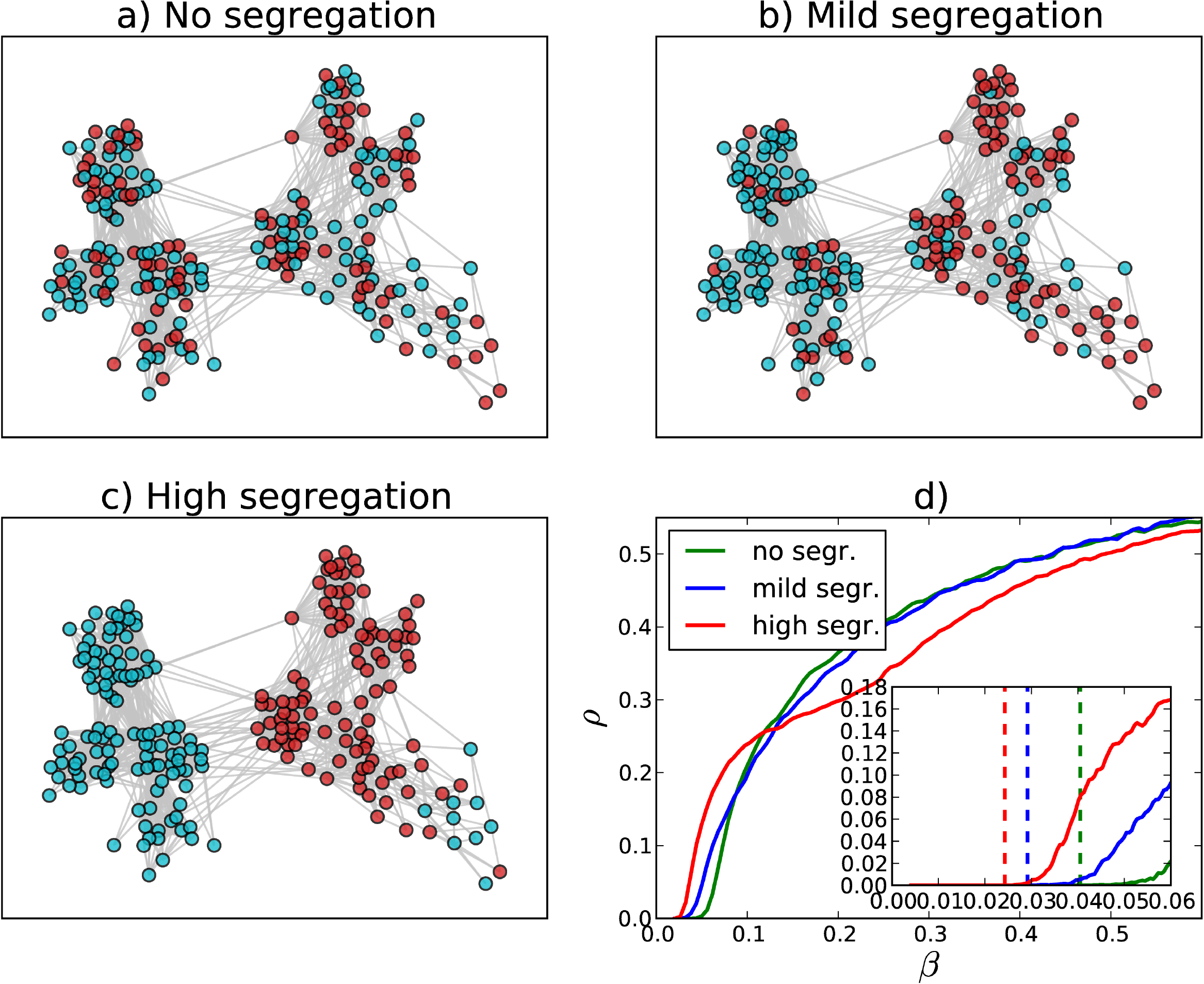}
\end{center} 
\caption{ {\bf The endemic state for different level of segregations and different susceptibility.} a), b) and c) face-to-face proximity school network with no, mild and high segregation respectively. High susceptibility nodes are colored red. d) The fraction of infected nodes in the endemic state averaged over 10,000 runs as a function of the susceptibility $\beta_1$ for the school network with three levels of segregation, along with the critical values of $\beta_1$ for which $R_0 = \lambda_{1,R} + 1 - \delta = 1$ (inset) denoted as horizontal dashed lines.}
\label{fig:viz}
\end{figure*}

Fig. \ref{fig:viz}a, 2b and 2c show the primary school network at three different iterations in the segregation process; no segregation, mild segregation and high segregation respectively, with high susceptibility nodes colored red. Fig. \ref{fig:viz}d shows the average fraction of infected nodes in the endemic state, $\rho$, averaged over $10,000$ simulations, for different values of $\beta_1$ for the three networks shown in Fig. \ref{fig:viz}a, b and c. As we mentioned earlier, and as can be seen in Fig. \ref{fig:viz}d, segregation can change the shape of the epidemic curve and depending on the network topology, high $R_0$ does not necessarily lead to a high number of infected nodes in the endemic state. This can happen when there is extreme segregation and the infection is contained within a small number of highly susceptibility nodes that have no or little contact to the rest of the network. In practice, however, we often observe mild segregation which increases both $R_0$ and the number of infected nodes in the endemic state. Finally, we compute the critical values of $\beta_1$ using equation (\ref{r0-qmf-general}). The inset of Fig. \ref{fig:viz}d shows $\rho$ near these critical values along with the critical values (horizontal dashed lines).

\section*{Discussion}
In this paper, we extended the SIS model to incorporate heterogenous susceptibility and showed that heterogeneity can significantly increase the networks' vulnerability to diseases. For individual level correlations between the susceptibility and the degree of a node, we find that approaches using the average susceptibility of the system to approximate $R_0$ will over- or underestimate the potential spread of diseases. In other words, when there is variation in susceptibility, the increase in susceptibility of a few individuals is not necessarily compensated by the decrease in susceptibility of others, since the degrees and locations of these individuals play an important role. On the other hand, we found that when nodes with similar characteristics are more likely to be connected, the vulnerability of the network increases, since a small group of densely connected high-susceptible individuals can act as a pocket supporting a persistent infection. Having a formula to compute $R_0$, however, has little practical use in absence of complete information about the network topology and the distribution of susceptibilities. In this case, we provided a method to approximate $R_0$ by coarse-graining the individual nodes into degree-susceptibility classes and only estimating the mixing patterns between these classes. Going forward, it is important that mathematical models of epidemic spreading include the effects of heterogenous susceptibility to provide more accurate descriptions of epidemic spreading processes. Otherwise, the basic reproductive numbers estimated from compartmental models will be systematically over or underestimated.


\section*{Acknowledgments}
L.K. thanks ONR Global, project ``Information fusion in networked sensors and systems'', for partial support. C.A.H. acknowledges the MIT Media Lab consortia and the ABC Career Development Chair.



\section*{Author contributions}
D.S., C.A.H. and L.K. designed the research. D.S. developed analytic tools and performed simulations. D.S., C.A.H. and L.K. wrote the paper, discussed results and reviewed the manuscript.

\section*{Competing financial interest}
The authors declare no competing financial interests.

\pagebreak

\section*{Figure Legends}

\end{document}